\documentclass[12pt]{article}
\usepackage{amssymb}
\usepackage[dvips]{epsfig}

\newcommand{\be}{\begin{equation}}
\newcommand{\ee}{\end{equation}}
\def\bea{\begin{eqnarray}}
\def\eea{\end{eqnarray}}
\newcommand{\bn}{\begin{eqnarray}}
\newcommand{\en}{\end{eqnarray}}
\newcommand{\nn}{\nonumber}
\newcommand{\no}{\noindent}

\newcommand{\p}{\partial}
\def\bea{\begin{eqnarray}}
\def\eea{\end{eqnarray}}
\newcommand{\beq}{\begin{eqnarray}}
\newcommand{\eeq}{\end{eqnarray}}

\begin{document}

\title{\textbf{More on dual actions for massive spin-2 particles}}
\author{D. Dalmazi$^{1}$\footnote{denis.dalmazi@unesp.br} and A. L. R. dos Santos$^{2}$\footnote{alessandroribeiros@yahoo.com.br} \\
\textit{{1- UNESP - Campus de Guaratinguet\'a - Departamento de F\'isica}}\\
\textit{{CEP 12516-410 - Guaratinguet\'a - SP - Brazil.} }\\
\textit{{2- Instituto Tecnol\'ogico de Aeron\'autica, DCTA} }\\
\textit{{CEP 12228-900, S\~ao Jos\'e dos Campos - SP - Brazil} }\\}
\date{\today}
\maketitle

\begin{abstract}

Here we start from a dual version of Vasiliev's first order action for massless spin-2 particles (linearized first order Einstein-Hilbert) and derive, via Kaluza-Klein dimensional reduction from $D+1$ to $D$ dimensions,
a set of dual massive spin-2 models. This set includes the  massive ``BR'' model, a spin-2 analogue of the spin-1 Cremmer-Scherk model.
In our approach the linearized Riemann curvature emerges from a solution of a functional constraint. In $D=2+1$ the BR model can be written as a linearized version of a new bimetric model for massive gravitons. We also have a new massive spin-2 model, in arbitrary dimensions, invariant under linearized diffeomorphisms. It is  given in terms of a non symmetric rank-2 tensor and a mixed symmetry tensor.

\end{abstract}

\newpage

\section{Introduction}

Although there is a very low ($ 10^{-23} \, eV$) experimental upper bound for the graviton mass, see \cite{gw2}, the only way to understand why is it massless (if it is) is to assume some nonzero mass and search for consequences. Earlier investigations \cite{vdv,zak} starting with  linearized massive gravitons (free massive spin-2 particles coupled to an external source) have found inconsistencies with either the Newtonian limit or the stars light deviation angle by the Sun (gravitational lens). The discrepancy persists no matter how close we are to the zero mass limit. This is the so called vDVZ mass discontinuity. Later, one has noticed that the graviton mass introduces another scale, the Vainshtein \cite{vain} radius ``$r_V$'', in the theory which increases as $m_{grav} \to 0$ thus, invalidating the linearized approximation which is supposed to hold beyond $r_V$.

The introduction of nonlinear terms, though may cure the vDVZ problem, lead in general to ghosts \cite{db}.
Such conflict has remained for many years without much progress
until the last decade when a clever choice  for a nonlinear graviton potential has appeared \cite{drgt} in order to tackle those problems.
Several works on massive gravity and bigravity \cite{hr} have followed \cite{drgt}, see the review works \cite{hinter,drham,sms}.

Massive gravity is an ongoing subject, there are still instability problems and an ultraviolet completion is needed. For our purposes it is important to notice that massive gravity and bigravity models require a graviton potential built up on the top of the usual free massive spin-2 Fierz-Pauli model \cite{fp} described by a symmetric rank two tensor. The use of dual spin-2 models described in terms of other tensors may inspire us to follow new routes in order to keep reparametrization symmetry without introducing Stueckelberg fields as in \cite{drgt} or another metric with propagating modes as in the bimetric model of \cite{hr}.

The purpose of the present work is to investigate massive spin-2 dual models by means of a dual parent action and present the known models and a new one in a unified way starting from the same parent action. The parent action method is reviewed in \cite{hl} in the context of the duality between massless $p$-forms and $(D-2-p)$-forms in $D$ dimensions. It is also used in \cite{bch} for obtaining dual models for massless  spin-s particles of arbitrary integer spins $s \ge 2$. Regarding the massive case, a first order parent action had appeared longo ago \cite{Curt} in the scalar tensor\footnote{This is the $(p,D)=(0,4)$ case of the duality between massive $p$-forms and $(D-p-1)$-forms} duality in $D=3+1$. In \cite{cmu} it is applied for massive spin-2 particles in $D=3+1$, see also the works \cite{zino03,zino09}.

Differently from those works, here we do not start with a massive parent action, we begin with a massless spin-2 action and perform a Kaluza-Klein dimensional reduction from $D+1$ to $D$ dimensions. We follow the approach  of \cite{k} and \cite{kmu,gkmu}. The infinite tower of massive modes do not mix at quadratic level so we can truncate the analysis to one massive mode. This is known to be consistent and preserve the number of degrees of freedom, see \cite{ady,rss}.
The gauge symmetries of the massless theory are inherited in general by the massive model through the presence of Stueckelberg fields. The profusion of fields generated by the dimensional reduction leads to a larger variety of dual models according to the set of fields that we choose to integrate over.

In the works \cite{bch,k,kmu,gkmu} a first order massless parent action of the Vasiliev type \cite{Vasiliev} is used. In the present work we use
a dual version of such action. This  allows us to derive a larger set of dual massive spin-2 models. In particular, a spin-2 version, see \cite{ks}, of the spin-1 topologically massive BF model or Cremmer-Scherk \cite{Cremmer} model  is now obtained. Differently from \cite{ks} where the linearized Riemann tensor is introduced in the dualization procedure in order to generate pure gauge solutions, here  the linearized Riemann tensor emerges naturally after solving a functional constraint. Those ideas might be useful to unravel the much less known higher spin geometries.

We also show here that in $D=2+1$ the model of \cite{ks}  can be interpreted as a linearized version of a new bimetric model very much inspired in the ``New Massive Gravity'' of \cite{bht}. Such model and its generalizations might be useful in discussing the interplay between bulk and boundary unitarity via a purely (bi)metric-formulation at action level, differently from \cite{mmg} and \cite{emg}.  Moreover, we obtain a new massive spin-2 model,  in arbitrary dimensions, described by a symmetric rank-2 tensor coupled to a mixed symmetry tensor.

We believe that our approach can be generalized for higher spins and curved backgrounds.

In the next section we introduce our basic procedure in the simpler case of
spin-1 particles in $D$ dimensions. In section 3 we deduce the dual massive parent action via KK dimensional reduction. We obtain the corresponding massive spin-2 dual models and present a new bimetric model for massive gravitons in $D=2+1$. In section 4 we draw our conclusions and perspectives.

\section{The spin-1 case}

In arbitrary $D$-dimensions\footnote{Throughout this work we use the metric
$\eta_{\mu\nu}=(-,+,+,\cdots,+)$ and the notation:
$e_{(\alpha\beta)}=(e_{\alpha\beta}+e_{\beta\alpha})/2$ and
$e_{[\alpha\beta]}=(e_{\alpha\beta}-e_{\beta\alpha})/2$. The symbols $S_P^{(s)}$ and $S_m^{(s)}$ stand for a parent action for massless and massive spin-s particle respectively.} the Maxwell theory can be written in a first order version with the help of an antisymmetric field $V_{[\mu\nu]}$,

\bea S_{P}^{(1)}=\int{d^{D}x}\Big[\,\frac{1}{4}V^{[\mu\nu]}V_{[\mu\nu]}-\frac{1}{2}V^{[\mu\nu]}(\partial_{\mu}A_{\nu}-\partial_{\nu}A_{\mu})\Big]\;.\label{Max-1a}\eea

\no The above parent action is invariant under $U(1)$ gauge transformations $\delta_{\varphi}{A}_{\mu}=\partial_{\mu}\varphi$. Integrating over the  auxiliary field $V^{[\mu\nu]}$ we recover the Maxwell theory; on the other hand the functional integral over the vector field furnishes the functional constraint $\partial_{\mu}V^{[\mu\nu]}=0$ whose general solution is $V^{[\mu\nu]}=\partial_{\alpha}T^{[\alpha\mu\nu]}$, where $T^{[\alpha\mu\nu]}$ is a completely antisymmetric but otherwise arbitrary tensor. Plugging back in (\ref{Max-1a}) we have  a second order model $S[T_{[\alpha\mu\nu]}]$ quadratic in $\partial_{\alpha}T^{[\alpha\mu\nu]}$ which on its turn can be written in a first order dual form:

\bea S_{P-Dual}^{(1)}=\int{d^{D}x}\Big[-\frac{1}{4}F^{[\mu\nu]}F_{[\mu\nu]}+\frac{1}{2}F_{[\mu\nu]}\partial_{\alpha}T^{[\alpha\mu\nu]}\Big]\;,\label{Tens-s1-1a}\eea

\no where now $F^{[\mu\nu]}$ and $T^{[\alpha\mu\nu]}$ are totally antisymmetric elementary and independent fields.
The dual parent action (\ref{Tens-s1-1a})
is invariant under  $\delta_{\Omega}{T}^{[\alpha\mu\nu]}=\partial_{\beta}\Omega^{[\beta\alpha\mu\nu]}$. Integrating over $F_{[\mu\nu]}$ returns $S[T_{[\alpha\mu\nu]}]$ while the integral over $T_{[\alpha\mu\nu]}$
leads to functional constraints of the Bianchi type:

\bea \partial^{\alpha}F^{[\mu\nu]}+
\partial^{\mu}F^{[\nu\alpha]}+\partial^{\nu}F^{[\alpha\mu]}=0\label{Bianchi-1}\;,\eea

\no whose general solution is $F^{[\mu\nu]}(A)=\partial^{\mu}A^{\nu}-\partial^{\nu}A^{\mu}$. Substituting $F^{[\mu\nu]}(A)$ in (\ref{Tens-s1-1a}) we recover the Maxwell theory. The Maxwell equations follow now from the $F_{[\mu\nu]}$ equations of motion:

\be F_{[\mu\nu]}= \p^{\alpha}T_{\alpha\mu\nu}\;. \label{fmneom} \ee

 \no Indeed,  (\ref{fmneom}) is equivalent to
 $\p^{\mu}F_{\mu\nu}=0$. Notice that in $D=3+1$ this is equivalent to
 $\epsilon^{\mu\nu\alpha\beta}\p_{\nu}\tilde{F}_{\alpha\beta} = 0 $ where $\tilde{F}^{\mu\nu}=\epsilon^{\mu\nu\alpha\beta}F_{\alpha\beta}$. So the equations of motion (\ref{Bianchi-1}) and (\ref{fmneom}) are equivalent to the explicitly eletric-magnetic duality invariant equations:

 \be dF = 0 \quad ; \qquad d\tilde{F} =0\;. \label{dualF} \ee

Next we perform the KK dimensional reduction of the dual  parent action (\ref{Tens-s1-1a}) in $D+1$ :

\bea S^{(1)}_{P-Dual}=\int{d^{D+1}x}\Big[-\frac{1}{4}F^{[MN]}F_{[MN]}+\frac{1}{2}F_{[MN]}\partial_{A}T^{[AMN]}\Big]\;,\label{s1-D+1}\eea

\no where we have introduced sources and used capital Latin letters to denote the $(D+1)$-dimensional indices, ($M,N,\ldots=0,1,2,\ldots,D$). In  $D$-dimensions we use Greek letters ($\mu,\nu,\ldots=0,1,2,\ldots,D-1$). We compactfy the last spatial dimension $x^{D}=y$ in a circle of radius $R=1/m$ and keep only one massive mode.  So the fields are redefined as

\bea F^{[MN]}(x^{\alpha},y)\rightarrow\left\{\begin{array}{l}
F^{[\mu\nu]}=\sqrt{\frac{m}{\pi}}\,F^{[\mu\nu]}(x)\cos{my}\\
F^{[\mu{D}]}=\sqrt{\frac{m}{\pi}}\,V^{\mu}(x)\sin{my}
\end{array}\right.\;,
\eea

\bea T^{[AMN]}(x^{\alpha},y)\rightarrow\left\{\begin{array}{l}
T^{[\alpha\mu\nu]}=\sqrt{\frac{m}{\pi}}\,T^{[\alpha\mu\nu]}(x)\cos{my}\\
T^{[\alpha\mu{D}]}=\sqrt{\frac{m}{\pi}}\,B^{[\alpha\mu]}(x)\sin{my}
\end{array}\right.\;.
\eea

\no Back in  (\ref{s1-D+1}) and integrating over $y$ we find the following
 $D$-dimensional massive parent action

\be S^{(1)}_m=\int{d^{D}x}\Big[-\frac{1}{4}F^{[\mu\nu]}F_{[\mu\nu]}+\frac{1}{2}F^{[\mu\nu]}\partial^{\alpha}T_{[\alpha\mu\nu]}-\frac{1}{2}V^{\mu}V_{\mu}+V^{\mu}\partial^{\nu}B_{[\nu\mu]}+\frac{m}{2}F^{[\mu\nu]}B_{[\mu\nu]}\Big],\label{Tens-s1-mass}\ee

\no which is invariant under the gauge transformations:

\bea  \delta_{\Phi}{B}_{[\mu\nu]}=\partial^{\beta}\Phi_{[\beta\mu\nu]}\qquad;\qquad \delta_{\Phi,\Omega}{T}_{[\alpha\mu\nu]}=-m\Phi_{[\alpha\mu\nu]}+\partial^{\beta}\Omega_{[\beta\alpha\mu\nu]}\;.\label{sym-dual-s1}\eea

\no Now we are able to deduce the known spin-1 massive dual models, namely, the $D$-dimensional version of the Kalb-Ramond \cite{Kalb}, Cremmer-Sherk \cite{Cremmer} and Maxwell-Proca \cite{Proca} models. By first integrating over $F^{[\mu\nu]}$ and $V^{\mu}$  we obtain from (\ref{Tens-s1-mass}) the Kalb-Ramond model with the Stueckelberg fields $T_{[\alpha\mu\nu]}$, namely
\bea
S_{KR}&=&\int{d^{D}x}\Big[\;\frac{1}{2}\partial_{\alpha}\widetilde{B}^{[\alpha\mu]}\partial^{\beta}\widetilde{B}_{[\beta\mu]}+\frac{m^{2}}{4}\widetilde{B}_{[\mu\nu]}\widetilde{B}^{[\mu\nu]}\Big]\;,\label{KR-Stuc}\eea

\no where $\widetilde{B}_{[\mu\nu]}=B_{[\mu\nu]}+\frac{\partial^{\alpha}T_{[\alpha\mu\nu]}}{m}$. The model is invariant under (\ref{sym-dual-s1}) and the fields $T^{[\alpha\mu\nu]}$ could be completely gauged away.

On the other hand, by first integrating over $T_{[\alpha\mu\nu]}$ in (\ref{Tens-s1-mass}) we obtain  the constraints (\ref{Bianchi-1}) and plugging the general solution $F^{[\mu\nu]}(A)=\partial^{\mu}A^{\nu}-\partial^{\nu}A^{\mu}$ back in the action we have the intermediate action:

\be S_{I}=\int{d^{D}x}\Big[-\frac{1}{4}F_{\mu\nu}(A)F^{\mu\nu}(A)-\frac{1}{2}V_{\mu}V^{\mu}-\frac{1}{2}B^{[\mu\nu]}F_{\mu\nu}(V-mA)\Big].\label{BF-1o}\ee

\no By further integrating over $V^{\mu}$ we arrive at the Cremmer-Scherk model in $D$-dimensions\footnote{In $D=3+1$ we can redefine $B^{\mu\nu}=\epsilon^{\mu\nu\alpha\beta}\widehat{B}_{\alpha\beta}$ in order to obtain the so called topologically massive BF model or simply Cremmer-Scherk model.}

\be S_{BF}=\int{d^{D}x}\Big[-\frac{1}{4}F_{\mu\nu}(A)F^{\mu\nu}(A)+\frac{1}{2}\partial_{\alpha}B^{[\alpha\mu]}\partial^{\beta}B_{[\beta\mu]}-\frac{m}{2}B^{[\mu\nu]}F_{\mu\nu}(A)\Big]\;.\label{BF}\ee

\no It is invariant under $\delta_{\Phi}{B}_{[\mu\nu]}=\partial^{\alpha}\Phi_{[\alpha\mu\nu]}$ and $\delta_{\varphi}{A}_{\mu}=\partial_{\mu}\varphi$.
It is a remarkable model, it describes massive spin-1 particles while preserving the $U(1)$ gauge symmetry without Stueckelberg fields. It can be generalized to the non-Abelian case \cite{ft}.

If instead of integrating over $V^{\mu}$ we integrate over $B^{[\mu\nu]}$ in (\ref{BF-1o})
we obtain the zero curvature constraint $\partial_{\mu}(V_{\nu}-mA_{\nu})-\partial_{\nu}(V_{\mu}-mA_{\mu})=0$  whose general solution $V_{\mu}=mA_{\mu}+\partial_{\mu}\psi$ leads to the
Maxwell-Proca theory

\bea S_{Proca}&=&\int{d^{D}x}\Big[-\frac{1}{4}F_{\mu\nu}(\widetilde{A})F^{\mu\nu}(\widetilde{A})-\frac{m^{2}}{2}\widetilde{A}^{\;2}_{\mu}\Big]\;.\label{Proca-Stuc}\eea

\no where $\widetilde{A}_{\mu}={A}_{\mu}+\frac{\partial_{\mu}\psi}{m}$. Now the gauge invariance is assured by the presence of Stueckelberg field $\psi$. All three models
$S_{KB}$, $S_{BF}$ and $S_{Proca}$ are gauge invariant as a consequence of the gauge symmetry of the massless parent action (4).
If we had started from the massive parent action which stems from the dimensional reduction of (\ref{Max-1a}) we would also have derived the three models $S_{KR}$, $S_{BF}$ and $S_{Proca}$, as in \cite{k}, but the role of the Gaussian integrals and the functional constraints would be interchanged.

\section{Dual massive spin-2 models}

Massless spin-2 particles are commonly described (in $D$ dimensions) by the linearized Einstein-Hilbert theory in terms of a symmetric rank-2 tensor field. This theory can be written in a first order version using a non-symmetric rank-2 tensor, $e_{\mu\nu}$, and a mixed symmetry rank-3 tensor, $Y^{\mu[\alpha\beta]}$, as \cite{Vasiliev,bch,Ajith}

\bea S^{(2)}_{P}&=&\frac{1}{2}\int{d^{D}x}\Big[Y^{\mu[\alpha\beta]}Y_{\beta[\alpha\mu]}-\frac{Y^{\alpha}Y_{\alpha}}{(D-2)}-Y^{\mu[\alpha\beta]}(\partial_{\alpha}e_{\beta\mu}-\partial_{\beta}e_{\alpha\mu})\Big]\;,\label{LEH-1a}\eea

\no where $Y^{\alpha}=\eta_{\mu\beta}Y^{\mu[\alpha\beta]}$. Integrating over the auxiliary field $Y^{\mu[\alpha\beta]}$ we recover the linearized Einstein-Hilbert theory in terms of $e_{(\mu\nu)}$. On the other hand the functional integral over $e_{\mu\nu}$ furnishes the functional constraint $\partial_{\alpha}Y^{\mu[\alpha\beta]}=0$ whose general solution is $Y^{\mu[\alpha\beta]}=\partial_{\nu}T^{\mu[\nu\alpha\beta]}$. Replacing this result back in (\ref{LEH-1a}) we have a second order model $S[T_{\mu[\nu\alpha\beta]}]$ quadratic in $\partial_{\nu}T^{\mu[\nu\alpha\beta]}$.

Similar to the linearized Einstein-Hilbert theory, the model $S[T_{\mu[\nu\alpha\beta]}]$ can also be written in a dual first order version (see \cite{bch})

\bea S^{(2)}_{P-Dual}&=&\int{d^{D}x}\Big[-\frac{1}{8}Y^{\mu[\alpha\beta]}Y_{\mu[\alpha\beta]}-\frac{1}{4}Y^{\mu[\alpha\beta]}Y_{\beta[\alpha\mu]}+\frac{1}{2}Y^{\alpha}Y_{\alpha}-\frac{1}{2}Y^{\mu[\alpha\beta]}\partial^{\nu}T_{\mu[\nu\alpha\beta]}\Big]\,,\label{Tens-s2-1a}\eea

\no which is invariant under vector gauge transformations
\bea \delta_{\xi}Y^{\mu[\alpha\beta]}&=&-\partial^{\mu}\partial^{\alpha}\xi^{\beta}+\partial^{\mu}\partial^{\beta}\xi^{\alpha}\;,\nn\\
\delta_{\xi}T^{\mu[\nu\alpha\beta]}&=&\eta^{\mu\nu}(\partial^{\alpha}\xi^{\beta}-\partial^{\beta}\xi^{\alpha})+\eta^{\mu\alpha}(\partial^{\beta}\xi^{\nu}-\partial^{\nu}\xi^{\beta})+\eta^{\mu\beta}(\partial^{\nu}\xi^{\alpha}-\partial^{\alpha}\xi^{\nu})\;,\eea
and antisymmetric shifts
\bea \delta_{\omega}Y^{\mu[\alpha\beta]}&=&-\partial^{\alpha}\omega^{[\beta\mu]}+\partial^{\beta}\omega^{[\alpha\mu]}\;,\nn\\
\delta_{\omega,\,\Omega}T^{\mu[\nu\alpha\beta]}&=&-\eta^{\mu\nu}\omega^{[\alpha\beta]}-\eta^{\mu\alpha}\omega^{[\beta\nu]}-\eta^{\mu\beta}\omega^{[\nu\alpha]}+\partial_{\sigma}\Omega^{\mu[\sigma\nu\alpha\beta]}\;,\eea
where $\xi_{\mu}$, $\omega^{[\mu\nu]}$ and $\Omega^{\mu[\sigma\nu\alpha\beta]}$ are gauge parameters.

Integrating over $Y^{\mu[\alpha\beta]}$ we recover $S[T_{\mu[\nu\alpha\beta]}]$ while the integral over $T_{\mu[\nu\alpha\beta]}$ leads to constraints of the Bianchi type:
\bea \partial^{\nu}Y^{\mu[\alpha\beta]}+\partial^{\alpha}Y^{\mu[\beta\nu]}+\partial^{\beta}Y^{\mu[\nu\alpha]}=0\label{eq-Y}\;,\eea
whose general solution is $Y^{\mu[\alpha\beta]}=\partial^{\alpha}e^{\beta\mu}-\partial^{\beta}e^{\alpha\mu}$, where $e^{\mu\nu}$ is a non-symmetric rank-2 tensor.  Plugging back in (\ref{Tens-s2-1a}) we have the linearized Einstein-Hilbert theory in terms of $e_{(\mu\nu)}$. In $D=3+1$ we can redefine $Y^{\mu[\alpha\beta]}=\epsilon^{\alpha\beta\nu\sigma}\widehat{Y}^{\mu}_{\;\;\;[\nu\sigma]}$ and $T_{\mu[\nu\alpha\beta]}=\epsilon_{\nu\alpha\beta\sigma}e^{\sigma}_{\;\;\mu}$ and show that (\ref{LEH-1a}) and (\ref{Tens-s2-1a}) coincide, similarly for (\ref{Max-1a}) and (\ref{Tens-s1-1a}) in the  spin-1 case. However, the massive parent actions in $D=3+1$ descending from
(\ref{LEH-1a}) and (\ref{Tens-s2-1a}) in $D=4+1$ are different since the self-duality of the massless parent actions only hold in $D=3+1$.

Now we are ready to perform the dimensional reduction of (\ref{Tens-s2-1a}) in $D+1$:

\bea S^{(2)}_{P-Dual}\!&\!=\!&\!\int{\!d^{D+1}x}\Big[\!-\frac{1}{8}Y^{M[AB]}Y_{M[AB]}-\frac{1}{4}Y^{M[AB]}Y_{B[AM]}+\frac{1}{2}Y^{A}Y_{A}-\frac{1}{2}Y^{M[AB]}\partial^{N}T_{M[NAB]}\Big].\nn\\
\label{s2-D+1}\eea
Compactfying the spatial dimension $y=x^{D}$ in a circle, the fields and the gauge parameters are redefined according to
\bea Y^{M[AB]}(x^{\alpha},y)&\rightarrow&\left\{\begin{array}{l}
Y^{\mu[\alpha\beta]}=\sqrt{\frac{m}{\pi}}\,Y^{\mu[\alpha\beta]}(x)\cos{my}\\
Y^{\mu[\alpha{D}]}=\sqrt{\frac{m}{\pi}}\,W^{\mu\alpha}(x)\sin{my}\\
Y^{D[\alpha\beta]}=\sqrt{\frac{m}{\pi}}\,V^{[\alpha\beta]}(x)\sin{my}\\
Y^{D[\alpha{D}]}=\sqrt{\frac{m}{\pi}}\,Z^{\alpha}(x)\cos{my}
\end{array}\right.\;,\label{w}\\
\nn\\
T^{M[NAB]}(x^{\alpha},y)&\rightarrow&\left\{\begin{array}{l}
T^{\mu[\nu\alpha\beta]}=\sqrt{\frac{m}{\pi}}\,T^{\mu[\nu\alpha\beta]}(x)\cos{my}\\
T^{\mu[\nu\alpha{D}]}=\sqrt{\frac{m}{\pi}}\,B^{\mu[\nu\alpha]}(x)\sin{my}\\
T^{D[\nu\alpha\beta]}=\sqrt{\frac{m}{\pi}}\,M^{[\nu\alpha\beta]}(x)\sin{my}\\
T^{D[\nu\alpha{D}]}=\sqrt{\frac{m}{\pi}}\,N^{[\nu\alpha]}(x)\cos{my}
\end{array}\right.\;,\label{x}\\
\nn\\
\xi^{M}(x^{\alpha},y)&\rightarrow&\left\{\begin{array}{l}
\xi^{\mu}=\sqrt{\frac{m}{\pi}}\xi^{\mu}(x)\cos{my}\\
\xi^{D}=\sqrt{\frac{m}{\pi}}\varphi(x)\sin{my}\\
\end{array}\right.\;,\label{y}\\
\nn\\
\omega^{[MN]}(x^{\alpha},y)&\rightarrow&\left\{\begin{array}{l}
\omega^{[\mu\nu]}=\sqrt{\frac{m}{\pi}}\Lambda^{[\mu\nu]}(x)\cos{my}\\
\omega^{[\mu{D}]}=\sqrt{\frac{m}{\pi}}\chi^{\mu}(x)\sin{my}\\
\end{array}\right.\;,\label{z}\\
\nn\\
\Omega^{M[CNAB]}(x^{\alpha},y)&\rightarrow&\left\{\begin{array}{l}
\Omega^{\mu[\sigma\nu\alpha\beta]}=\sqrt{\frac{m}{\pi}}\,\Omega^{\mu[\sigma\nu\alpha\beta]}(x)\cos{my}\\
\Omega^{\mu[\sigma\nu\alpha{D}]}=\sqrt{\frac{m}{\pi}}\,\Theta^{\mu[\sigma\nu\alpha]}(x)\sin{my}\\
\Omega^{D[\sigma\nu\alpha\beta]}=\sqrt{\frac{m}{\pi}}\,\Gamma^{[\sigma\nu\alpha\beta]}(x)\sin{my}\\
\Omega^{D[\sigma\nu\alpha{D}]}=\sqrt{\frac{m}{\pi}}\,\lambda^{[\sigma\nu\alpha]}(x)\cos{my}
\end{array}\right.\;,\label{zz}\eea

\no where $W_{\mu\nu}$ is an arbitrary rank-2 tensor without symmetry. Substituting (\ref{w}-\ref{x}) in (\ref{s2-D+1}) and integrating over $y$ we obtain a first-order massive spin-2 parent action given by

\bea S^{(2)}_{m}\!&\!=\!&\!\int{\!d^{D}x}\Bigg\{\!-\frac{1}{8}Y^{\mu[\alpha\beta]}Y_{\mu[\alpha\beta]}-\frac{1}{4}Y^{\mu[\alpha\beta]}Y_{\beta[\alpha\mu]}+\frac{1}{2}Y^{\alpha}Y_{\alpha}-\frac{1}{2}Y^{\mu[\alpha\beta]}[\partial^{\nu}T_{\mu[\nu\alpha\beta]}+m\,B_{\mu[\alpha\beta]}]\nn\\
&&\qquad\qquad+Y^{\mu}Z_{\mu}-\frac{1}{2}W^{(\mu\nu)}W_{(\mu\nu)}+\frac{1}{2}W^{2}+W^{\mu\nu}\partial^{\alpha}B_{\mu[\nu\alpha]}+\frac{1}{2}W^{[\mu\nu]}V_{[\mu\nu]}-\frac{1}{8}V^{[\mu\nu]}V_{[\mu\nu]}\nn\\
&&\qquad\qquad+\frac{m}{2}V^{[\mu\nu]}N_{[\mu\nu]}-\frac{1}{2}V^{[\mu\nu]}\partial^{\alpha}M_{[\alpha\mu\nu]}+Z^{\mu}\partial^{\nu}N_{[\mu\nu]}\Bigg\}\;,\label{Tens-s2-mass}\eea

\no which is invariant under the following independent transformations

\bea
\delta_{\xi}Y^{\mu[\alpha\beta]}&=&-\partial^{\mu}\partial^{\alpha}\xi^{\beta}+\partial^{\mu}\partial^{\beta}\xi^{\alpha}\qquad;\qquad
\delta_{\xi,\,\varphi}W^{\mu\nu}=-m\partial^{\mu}\xi^{\nu}-\partial^{\mu}\partial^{\nu}\varphi\nn\\
\delta_{\xi}V^{[\alpha\beta]}&=&m(\partial^{\alpha}\xi^{\beta}-\partial^{\beta}\xi^{\alpha})\qquad\quad;\qquad \delta_{\xi,\,\varphi}Z^{\mu}=-m^{2}\xi^{\mu}-m\partial^{\mu}\varphi\nn\\
\delta_{\xi}T^{\mu[\nu\alpha\beta]}&=&\eta^{\mu\nu}(\partial^{\alpha}\xi^{\beta}-\partial^{\beta}\xi^{\alpha})+\eta^{\mu\alpha}(\partial^{\beta}\xi^{\nu}-\partial^{\nu}\xi^{\beta})+\eta^{\mu\beta}(\partial^{\nu}\xi^{\alpha}-\partial^{\alpha}\xi^{\nu})\nn\\
\delta_{\xi,\,\varphi}B^{\mu[\alpha\beta]}&=&\eta^{\mu\alpha}(m\xi^{\beta}+\partial^{\beta}\varphi)-\eta^{\mu\beta}(m\xi^{\alpha}+\partial^{\alpha}\varphi)\nn\\
\delta_{\xi}N^{[\alpha\beta]}&=&\partial^{\alpha}\xi^{\beta}-\partial^{\beta}\xi^{\alpha}\qquad\qquad\quad;\qquad\delta_{\xi,\,\varphi}{M}^{[\nu\alpha\beta]}=0\label{tr-1-m}\eea
and
\bea \delta_{\Lambda}Y^{\mu[\alpha\beta]}&=&-\partial^{\alpha}\Lambda^{[\beta\mu]}+\partial^{\beta}\Lambda^{[\alpha\mu]}\qquad;\qquad
\delta_{\Lambda,\,\chi}W^{\mu\nu}=m\Lambda^{[\mu\nu]}+\partial^{\nu}\chi^{\mu}\nn\\
\delta_{\chi}V^{[\alpha\beta]}&=&-\partial^{\alpha}\chi^{\beta}+\partial^{\beta}\chi^{\alpha}\qquad\qquad;\qquad \delta_{\chi}Z^{\mu}=m\chi^{\mu}\nn\\
\delta_{\Lambda,\,\Omega,\,\Theta}T^{\mu[\nu\alpha\beta]}&=&-\eta^{\mu\nu}\Lambda^{[\alpha\beta]}-\eta^{\mu\alpha}\Lambda^{[\beta\nu]}-\eta^{\mu\beta}\Lambda^{[\nu\alpha]}+\p_{\sigma}\Omega^{\mu[\sigma\nu\alpha\beta]}-m\Theta^{\mu[\nu\alpha\beta]}\nn\\
\delta_{\chi,\,\Theta}B^{\mu[\alpha\beta]}&=&-\eta^{\mu\alpha}\chi^{\beta}+\eta^{\mu\beta}\chi^{\alpha}+\partial_{\sigma}\Theta^{\mu[\sigma\alpha\beta]}\nn\\
\delta_{\Lambda,\,\lambda}N^{[\alpha\beta]}&=&-\Lambda^{[\alpha\beta]}+\p_{\sigma}\lambda^{[\sigma\alpha\beta]}\qquad\;;\qquad \delta_{\lambda,\,\Gamma}{M}^{[\nu\alpha\beta]}= m\,\lambda^{[\nu\alpha\beta]}+\partial_{\sigma}\Gamma^{[\sigma\nu\alpha\beta]}\label{tr-2-m}\;.\eea

\no Now we are ready to derive a set of dual massive spin-2 models.

\subsection{The Curtright-Freund and Fierz-Pauli models ($m\neq0$)}

The Fierz-Pauli (FP) \cite{fp}, Curtright-Freund (CF) \cite{Curt} and Cassini-Montemayor-Urrutia (CMU) \cite{cmu} models describe massive spin-2 particles. The first one in terms of a symmetric rank-2 tensor while the second and third ones in terms of a mixed symmetry rank-3 tensor. In order to obtain the CF model in $D$ dimensions from the massive parent action (\ref{Tens-s2-mass}) we start by integrating over the field $Y^{\mu[\alpha\beta]}$ followed by an integration over the symmetric part of $W^{\mu\nu}$. This leads us to the following intermediate action

\bea S_{II}&=&\int{d^{D}x}\Bigg\{\,\frac{1}{8}\Big(\partial_{\alpha}B^{\mu[\alpha\beta]}+\partial_{\alpha}B^{\beta[\alpha\mu]}\Big)^{2}-\frac{(\partial_{\alpha}B^{\alpha})^{2}}{2(D-1)}-W^{[\mu\nu]}\partial^{\alpha}B_{\mu[\alpha\nu]}+\frac{1}{2}W^{[\mu\nu]}V_{[\mu\nu]}\nn\\
&&\qquad\quad-\frac{1}{8}V^{[\mu\nu]}V_{[\mu\nu]}-\frac{1}{2}V^{[\mu\nu]}\partial^{\alpha}M_{[\alpha\mu\nu]}+\frac{m}{2}V^{[\mu\nu]}N_{[\mu\nu]}-Z^{\mu}\partial^{\nu}N_{[\nu\mu]}\nn\\
&&\qquad\quad+\frac{m^{2}}{2}\Big[B^{\mu[\alpha\beta]}+\frac{\partial_{\lambda}T^{\mu[\lambda\alpha\beta]}}{m}+\frac{1}{m}(\eta^{\mu\alpha}Z^{\beta}-\eta^{\mu\beta}Z^{\alpha})\Big]\nn\\
&&\qquad\qquad\quad\cdot\Big[B_{\alpha[\mu\beta]}+\frac{\partial^{\nu}T_{\alpha[\nu\mu\beta]}}{m}+\frac{1}{m}(\eta_{\alpha\mu}Z_{\beta}-\eta_{\alpha\beta}Z_{\mu})\Big]\nn\\
&&\qquad\quad-\frac{m^{2}}{2(D-2)}\Big[B^{\mu}+\frac{\partial_{\nu}T^{[\nu\mu]}}{m}-\frac{(D-1)Z^{\mu}}{m}\Big]^{2}\Bigg\}\;,\label{CF-01}\eea

\no where $B^{\alpha}=\eta_{\mu\beta}B^{\mu[\alpha\beta]}$ and $T^{[\alpha\beta]}=\eta_{\mu\nu}T^{\mu[\alpha\beta\nu]}$. The functional integration over $W_{[\mu\nu]}$ results in $V^{[\mu\nu]}=\partial_{\alpha}B^{\mu[\alpha\nu]}-\partial_{\alpha}B^{\nu[\alpha\mu]}$. Substituting $V^{[\mu\nu]}$ in (\ref{CF-01}) we obtain the CF model with Stueckelberg fields

\bea S_{CF}&\!=\!&\int{d^{D}x}\Bigg\{\,\frac{1}{2}\partial_{\alpha}\widetilde{B}^{\mu[\alpha\beta]}\partial^{\nu}\widetilde{B}_{\beta[\nu\mu]}-\frac{(\partial_{\alpha}\widetilde{B}^{\alpha})^{2}}{2(D-1)}+\frac{m^{2}}{2}\Big[\widetilde{B}^{\mu[\alpha\beta]}\widetilde{B}_{\alpha[\mu\beta]}-\frac{\widetilde{B}_{\mu}^{\;2}}{(D-2)}\Big]\Bigg\}\;,\label{CF}\eea

\no where

\bea \widetilde{B}^{\mu[\alpha\beta]}&=&B^{\mu[\alpha\beta]}+\frac{\partial_{\nu}T^{\mu[\nu\alpha\beta]}}{m}+\frac{1}{m}\eta^{\mu\alpha}(Z^{\beta}+\partial_{\nu}N^{[\nu\beta]})-\frac{1}{m}\eta^{\mu\beta}(Z^{\alpha}+\partial_{\nu}N^{[\nu\alpha]})\nn\\
&&-\frac{1}{m}\partial^{\mu}\Big(N^{[\alpha\beta]}-\frac{\partial_{\nu}M^{[\nu\alpha\beta]}}{m}\Big)\;.\eea

\no Due to the presence of the Stueckelberg fields, the CF action is invariant under the transformations (\ref{tr-1-m}) and (\ref{tr-2-m}). Furthermore, we can decompose $\widetilde{B}^{\mu[\alpha\beta]}$ in terms of its traceless piece $\overline{B}^{\;\mu[\alpha\beta]}$ and its trace $\widetilde{B}^{\alpha}$ as
\bea \widetilde{B}^{\mu[\alpha\beta]}=\overline{B}^{\;\mu[\alpha\beta]}+\frac{1}{(D-1)}(\eta^{\mu\beta}\widetilde{B}^{\alpha}-\eta^{\mu\alpha}\widetilde{B}^{\beta})\;.\eea

\no Back in (\ref{CF}) and neglecting the
trace $\widetilde{B}^{\alpha}$ which decouples, we have the compact action for massive spin-2 particles \cite{kmu,gkmu}:
\bea S_{CF}&\!=\!&\int{d^{D}x}\Big\{\,\frac{1}{2}\partial_{\alpha}\overline{B}^{\;\mu[\alpha\beta]}\partial^{\nu}\overline{B}_{\beta[\nu\mu]}+\frac{m^{2}}{2}\overline{B}^{\;\mu[\beta\alpha]}\overline{B}_{\alpha[\beta\mu]}\Big\}\;,\label{CFT}\eea

\no which coincides with the CF action \cite{Curt} as
shown in (\ref{CFT}).\footnote{We can also start from (\ref{CF-01})
and integrate first over $Z_{\mu}$ and then over $W_{[\mu\nu]}$ which allows us to eliminate $V_{[\mu\nu]}$. We end up with the flat space limit of the model of \cite{zino4} in terms of the mixed symmetry tensor $ \widetilde{X}^{\mu[\alpha\beta]}=X^{\mu[\alpha\beta]}+\partial_{\nu}T^{\mu[\nu\alpha\beta]}/m$ and the antisymmetric field $F^{[\mu\nu]}= N^{[\mu\nu]}-\partial_{\alpha}M^{[\alpha\mu\nu]}/m $. The same model appears in the more recent work \cite{bcc} where dual versions of massive, massless and partially massless spin-2 models  are discussed.}

On the other hand, in order to obtain the FP model from action (\ref{Tens-s2-mass}) we start integrating over $V^{[\mu\nu]}$ followed by an integration over $T_{\mu[\nu\alpha\beta]}$. The second integration results in equation (\ref{eq-Y}), whose general solution is $Y^{\mu[\alpha\beta]}=\partial^{\alpha}e^{\beta\mu}-\partial^{\beta}e^{\alpha\mu}$ where $e_{\mu\nu}$ is an arbitrary rank-2 tensor. We arrive at the intermediate action

\bea S_{III}&=&\int{d^{D}x}\Bigg\{-\frac{1}{2}\partial_{\mu}e_{(\alpha\beta)}\partial^{\mu}e^{(\alpha\beta)}+\frac{1}{2}\partial_{\mu}e\partial^{\mu}e-\partial_{\mu}e\partial_{\nu}e^{\mu\nu}+\partial_{\mu}e^{(\mu\nu)}\partial^{\alpha}e_{(\alpha\nu)}\nn\\
&&\qquad+\frac{1}{2}\Big[B^{\mu[\alpha\beta]}+\frac{1}{m}(\eta^{\mu\alpha}Z^{\beta}-\eta^{\mu\beta}Z^{\alpha})\Big]\cdot\Big[\partial_{\alpha}(W_{\mu\beta}-me_{\beta\mu})-\partial_{\beta}(W_{\mu\alpha}-me_{\alpha\mu})\Big]\nn\\
&&\qquad-\frac{1}{2}\Big[W^{\mu\nu}+mN^{[\mu\nu]}-\frac{\partial^{\nu}Z^{\mu}}{m}-\partial_{\beta}M^{[\beta\mu\nu]}\Big]\cdot\Big[W_{\nu\mu}+mN_{[\nu\mu]}-\frac{\partial_{\mu}Z_{\nu}}{m}-\partial^{\alpha}M_{[\alpha\nu\mu]}\Big]\nn\\
&&\qquad+\frac{1}{2}\Big[W-\frac{\partial^{\alpha}Z_{\alpha}}{m}\Big]^{2}\Bigg\}\;.\label{BT-BR}\eea

The functional integration over $B^{\mu[\alpha\beta]}$ furnish the functional constraint
\bea \partial_{\alpha}(W_{\mu\beta}-me_{\beta\mu})-\partial_{\beta}(W_{\mu\alpha}-me_{\alpha\mu})=0\;,\eea

\no whose general solution is $W_{\mu\beta}=me_{\beta\mu}+\partial_{\beta}A_{\mu}$, where $A_{\mu}$ is an arbitrary vector field. Putting back in (\ref{BT-BR}) we have the FP model

\bea S_{FP}&=&\int{d^{D}x}\Big\{-\frac{1}{2}\partial_{\mu}\tilde{e}_{(\alpha\beta)}\partial^{\mu}\tilde{e}^{(\alpha\beta)}+\frac{1}{2}\partial_{\mu}\tilde{e}\partial^{\mu}\tilde{e}-\partial_{\mu}\tilde{e}\partial_{\nu}\tilde{e}^{\mu\nu}+\partial_{\mu}\tilde{e}^{(\mu\nu)}\partial^{\alpha}\tilde{e}_{(\alpha\nu)}\nn\\
&&\qquad\quad-\frac{m^{2}}{2}[\tilde{e}_{\mu\nu}\tilde{e}^{\;\nu\mu}-\tilde{e}^{\;2}]\Big\}\;,\label{FP}\eea

\no where

\bea \tilde{e}_{\mu\nu}=e_{\mu\nu}+\frac{1}{m}\partial_{\mu}\Big(A_{\nu}-\frac{Z_{\nu}}{m}\Big)-N_{[\mu\nu]}+\frac{\partial^{\beta}M_{[\beta\mu\nu]}}{m}\;.\eea

\no $\tilde{e}_{\mu\nu}$ is invariant under the transformation (\ref{tr-1-m}) and (\ref{tr-2-m}) altogether with

\bea \delta_{\xi,\,\Lambda}e_{\mu\nu}&=&-\partial_{\nu}\xi_{\mu}-\Lambda_{[\mu\nu]}\quad;\qquad \delta_{\varphi,\,\chi}A_{\mu}=-\partial_{\mu}\varphi+\chi_{\mu}\label{tr-e-A}\;.\eea

\no The antisymmetric part $\tilde{e}_{[\mu\nu]}$ decouples from the symmetric one and vanishes on shell.

From (\ref{tr-2-m}) we see that we can use the gauge parameters $\Lambda_{[\mu\nu]}$, $\xi_{\mu}$ and $\lambda_{[\mu\nu\alpha]}$ to fix the gauges  $W_{[\mu\nu]}=0=Z_{\mu}=M_{[\alpha\mu\nu ]}$ in (\ref{CF-01}). Consequently,
$V_{[\mu\nu]}$ and  $N_{[\mu\nu]}$ decouple and can be integrated away. We end up with the CMU model \cite{cmu} in terms of the combination $B_{\mu[\alpha\beta]} + \p^{\nu}T_{\mu[\nu\alpha\beta]}/m $.

\subsection{The new massive spin-2 model}

If we integrate over the field $W_{\mu\nu}$ in (\ref{BT-BR}) we find a new massive spin-2 model

\bea S[\hat{e},\widehat{B}]&=&\int{d^{D}x}\Big\{-\frac{1}{2}\partial_{\mu}\hat{e}_{(\alpha\beta)}\partial^{\mu}\hat{e}^{(\alpha\beta)}+\frac{1}{2}\partial_{\mu}\hat{e}\partial^{\mu}\hat{e}-\partial_{\mu}\hat{e}\partial_{\nu}\hat{e}^{\mu\nu}+\partial_{\mu}\hat{e}^{(\mu\nu)}\partial^{\alpha}\hat{e}_{(\alpha\nu)}\nn\\
&&\qquad-\frac{m}{2}\widehat{B}^{\mu[\alpha\beta]}(\partial_{\alpha}\hat{e}_{\beta\mu}-\partial_{\beta}\hat{e}_{\alpha\mu})+\frac{1}{2}\partial_{\alpha}\widehat{B}^{\mu[\alpha\nu]}\partial^{\beta}\widehat{B}_{\nu[\beta\mu]}-\frac{\partial_{\alpha}\widehat{B}^{\alpha}\partial^{\beta}\widehat{B}_{\beta}}{2(D-1)}\Bigg\}\;,\label{BT}\eea

\no where
\bea \hat{e}_{\mu\nu}&=&e_{\mu\nu}-N_{[\mu\nu]}+\frac{1}{m}\partial^{\alpha}M_{[\alpha\mu\nu]}\;,\\
\widehat{B}^{\mu[\alpha\beta]}&=&B^{\mu[\alpha\beta]}+\frac{1}{m}\Big(\eta^{\mu\alpha}Z^{\beta}-\eta^{\mu\beta}Z^{\alpha}\Big)\;.\eea

Defining the tensor $F_{[\alpha\beta]\mu}\equiv\partial_{\alpha}\hat{e}_{\beta\mu}-\partial_{\beta}\hat{e}_{\alpha\mu}$ we can rewrite this new model as

\bea S[\hat{e},\widehat{B}]&=&\int{d^{D}x}\Big\{-\frac{1}{8}F^{[\alpha\beta]\mu}F_{[\alpha\beta]\mu}-\frac{1}{4}F^{[\alpha\beta]\mu}F_{[\alpha\mu]\alpha}+\frac{1}{2}F^{\alpha}F_{\alpha}-\frac{m}{2}\widehat{B}^{\mu[\alpha\beta]}F_{[\alpha\beta]\mu}\nn\\
&&\qquad\quad+\frac{1}{2}\partial_{\alpha}\widehat{B}^{\mu[\alpha\nu]}\partial^{\beta}\widehat{B}_{\nu[\beta\mu]}-\frac{\partial_{\alpha}\widehat{B}^{\alpha}\partial^{\beta}\widehat{B}_{\beta}}{2(D-1)}\Bigg\}\;,\label{Bw}\eea

\no where $F^{\alpha}=\eta_{\mu\beta}F^{[\alpha\beta]\mu}$. We can fix the unitary gauge $N_{[\alpha\beta]}=0=M_{[\alpha\mu\nu]}=Z^{\mu}$ and get rid of all remaining Stückelberg fields.  Alternatively we might choose $Z^{\mu}=m\,B^{\mu}/(D-1)$ such that $\eta_{\mu\beta}\widehat{B}^{\mu[\alpha\beta]}=\widehat{B}^{\alpha}=0$. Nevertheless, the model remain invariant under the independent gauge transformations:

\bea \delta_{\xi}\hat{e}_{\mu\nu}=-\partial_{\mu}\xi_{\nu}\quad;\qquad \delta_{\Theta}\widehat{B}^{\mu[\alpha\beta]}=\partial_{\sigma}\Theta^{\mu[\sigma\alpha\beta]}\;.\label{tr-e-X}\eea

\no The tensor $F_{[\alpha\beta]\mu}=\partial_{\alpha}\hat{e}_{\beta\mu}-\partial_{\beta}\hat{e}_{\alpha\mu}$ is some sort of linearized spin-connection.

The equations of motion from (\ref{Bw}) are given by:
\bea \partial_{\alpha}F^{[\alpha\beta]\mu}+\partial_{\alpha}F^{[\alpha\mu]\beta}+\partial_{\alpha}F^{[\mu\beta]\alpha}-2\eta^{\mu\beta}\partial_{\alpha}F^{\alpha}+2\partial^{\mu}F^{\beta}=-2m\partial_{\alpha}\widehat{B}^{\mu[\alpha\beta]}\;,\label{eq-1-Bw}\eea

\bea \partial^{\beta}\partial_{\nu}\widehat{B}^{\alpha[\nu\mu]}-\partial^{\alpha}\partial_{\nu}\widehat{B}^{\beta[\nu\mu]}+\frac{\eta^{\mu\beta}\partial^{\alpha}\partial_{\nu}\widehat{B}^{\nu}}{(D-1)}-\frac{\eta^{\mu\alpha}\partial^{\beta}\partial_{\nu}\widehat{B}^{\nu}}{(D-1)}=m\,F^{[\alpha\beta]\mu}\;.\label{eq-2-Bw}\eea

\no Applying $\eta_{\mu\beta}$ and $\partial_{\mu}$ on (\ref{eq-2-Bw}) we obtain respectively
\bea F^{\alpha}=0\qquad,\qquad \partial_{\mu}F^{[\alpha\beta]\mu}=0\;.\label{eq-3-Bw}\eea
On the other hand, applying $\eta_{\mu\beta}$ and $\partial_{\mu}$ on (\ref{eq-1-Bw}) and using the results (\ref{eq-3-Bw}) we obtain respectively
\bea \partial_{\alpha}\widehat{B}^{\alpha}=0\qquad,\qquad\partial_{\mu}\partial_{\alpha}\widehat{B}^{\mu[\alpha\beta]}=0\;.\label{eq-4-Bw}\eea
Substituting (\ref{eq-2-Bw}) in (\ref{eq-1-Bw}) and using the results (\ref{eq-3-Bw}-\ref{eq-4-Bw}) we find the Fierz-Pauli constraints and the Klein-Gordon equation,

\be \eta^{\mu\nu}\widehat{\mathbb{B}}_{(\mu\nu)} = 0 \quad , \quad \p^{\mu} \widehat{\mathbb{B}}_{(\mu\nu)} = 0 \quad , \quad \widehat{\mathbb{B}}_{[\mu\nu]} = 0\;, \label{fpc} \ee

\be (\square-m^{2})\widehat{\mathbb{B}}_{(\mu\nu)}=0\label{eq-6-Bw}\;.\ee

\no where we have defined $\widehat{\mathbb{B}}^{\mu\nu}\equiv\partial_{\alpha}\widehat{B}^{\mu[\alpha\nu]}$. This is all we need to describe massive spin-2 particles.
Notice that (\ref{eq-2-Bw}) gives $F_{[\alpha\beta]\mu}$, which are the relevant gauge invariants built up from $e_{\mu\nu}$, in terms of $\widehat{\mathbb{B}}_{(\mu\nu)}$. Had we introduced sources from the beginning we would have found the following dual map
between the Fierz-Pauli theory and the new model $S[\hat{e},\widehat{B}]$,

\bea e^{\mu\nu}\;\leftrightarrow\;\widehat{\mathbb{B}}^{\mu\nu}\;. \label{dmap}\eea

Regarding the massless limit ($m\rightarrow0$) of the new massive spin-2 model (\ref{BT}), it turns out that the Stueckelberg fields provide a smooth limit. Taking $m\rightarrow0$  in (\ref{BT}) we obtain
\bea S_{m=0}\!&\!=\!&\!\int{d^{D}x}\Big\{\!-\frac{1}{2}\partial_{\mu}e_{(\alpha\beta)}\partial^{\mu}e^{(\alpha\beta)}+\frac{1}{2}(\partial_{\mu}e)^{2}-\partial_{\mu}e\partial_{\nu}e^{\mu\nu}+(\partial_{\mu}e^{(\mu\nu)})^{2}+e_{\mu\nu}\partial^{\nu}Z^{\mu}-e\,\partial_{\mu}Z^{\mu}\nn\\
&&\qquad\qquad-N_{[\mu\nu]}\partial^{\mu}Z^{\nu}+M_{[\mu\nu\alpha]}\partial^{\mu}C^{[\nu\alpha]}+\frac{1}{2}C^{\mu\nu}C_{\nu\mu}-\frac{C^{2}}{2(D-1)}\Big\}\;,\label{new}\eea
where we have defined $C^{\mu\nu}=\partial_{\alpha}B^{\mu[\alpha\nu]}$. The integration over $N_{[\mu\nu]}$ leads to $Z^{\mu}=\partial^{\mu}\pi$ for some scalar field $\pi$, while the integration over $M_{[\mu\nu\alpha]}$ results in the Bianch identity $\partial^{\mu}C^{[\nu\alpha]}+\partial^{\nu}C^{[\alpha\mu]}+\partial^{\alpha}C^{[\mu\nu]}=0$ whose general solution is:

\bea C^{[\mu\nu]}&=& \partial_{\alpha}B^{\mu[\alpha\nu]}-\partial_{\alpha}B^{\nu[\alpha\mu]}=\partial^{\mu}A^{\nu}-\partial^{\nu}A^{\mu}\;, \label{gs1} \\
&&\nn\\
&\rightarrow & B^{\mu[\alpha\nu]} = \eta^{\mu\alpha}A^{\nu} -  \eta^{\mu\nu}A^{\alpha} + \p_{\gamma}B^{[\gamma\mu][\alpha\nu]} + \p_{\gamma} X^{\mu[\gamma\alpha\nu]} \; , \label{gs2} \eea

\no where $B^{[\beta\mu][\alpha\nu]}$ has the same index structure of the
Riemann curvature tensor while $X^{\mu[\gamma\alpha\nu]}$ is fully antisymmetric in the last three indices but otherwise arbitrary. Substituting back in (\ref{new}) and making the field redefinition $e_{\mu\nu}=e'_{\mu\nu}-\pi\eta_{\mu\nu}/(D-2)$ we find

\bea S_{m=0}\!&\!=\!&\!\int{d^{D}x}\Big\{\!-\frac{1}{2}\partial_{\mu}e'_{(\alpha\beta)}\partial^{\mu}e'^{(\alpha\beta)}+\frac{1}{2}(\partial_{\mu}e')^{2}-\partial_{\mu}e'\partial_{\nu}e'^{\mu\nu}+(\partial_{\mu}e'^{(\mu\nu)})^{2}\nn\\
&&\qquad\qquad+\frac{(D-1)}{2(D-2)}\pi\square\pi+\frac{1}{2}(\partial_{\alpha}\partial_{\beta}B^{[\alpha\mu][\beta\nu]})^{2}-\frac{1}{2}\frac{(\partial_{\alpha}\partial_{\beta}B^{\alpha\beta})^{2}}{(D-1)}\Big\}\;.\label{m=0}\eea

\no The last two terms in (\ref{m=0}) are equivalent to the Maxwell theory as shown in \cite{Siegel}. Therefore, the number of degrees of freedom of the massive case is preserved in the massless limit. In particular, in $D=3+1$ there are manifestly 5 degrees of freedom, two in the massless  tensor $e'_{(\mu\nu)}$, two in the higher rank electromagnetic field $B^{[\alpha\mu][\beta\nu]}$ and one in the massless scalar $\pi$.

\subsection{The massive BR model}

In \cite{ks} a fourth-order massive spin-2 model was proposed, with a BR-type coupling, where $B^{[\mu\alpha][\nu\beta]}$ is a auxiliary field and $R^{(1)}_{\mu\alpha\nu\beta}$ is the linearized Riemann curvature. Here we show that the BR model can be obtained from the massive parent action (\ref{Tens-s2-mass}), more specifically from its follow up action (\ref{BT-BR}). Integrating over the antisymmetric tensor $e_{\mu\nu}$ in (\ref{BT-BR}) we have the constraint

\bea \partial_{\alpha}\left(\widehat{B}^{\mu[\alpha\nu]}-\widehat{B}^{\nu[\alpha\mu]} \right)=0 \;, \label{eom1}\eea

\no whose general solution is given in (\ref{gs2}) with $A_{\mu}=0$.
 Substituting this result back in (\ref{BT-BR}) we find the second-order version of the BR model

\bea S^{(2)}_{BR}&=&\int{d^{D}x}\Bigg\{-\frac{1}{2}\partial_{\mu}e_{(\alpha\beta)}\partial^{\mu}e^{(\alpha\beta)}+\frac{1}{2}\partial_{\mu}e\partial^{\mu}e-\partial_{\mu}e\partial_{\nu}e^{\mu\nu}+\partial_{\mu}e^{(\mu\nu)}\partial^{\alpha}e_{(\alpha\nu)}\nn\\
&&\qquad-\frac{1}{2}\Big[W_{\mu\nu}-\frac{\partial_{\nu}Z_{\mu}}{m}+m\,N_{[\mu\nu]}-\partial^{\beta}M_{[\beta\mu\nu]}\Big].\Big[W^{\nu\mu}-\frac{\partial^{\mu}Z^{\nu}}{m}+m\,N^{[\nu\mu]}-\partial_{\alpha}M^{[\alpha\nu\mu]}\Big]\nn\\
&&\qquad+\frac{1}{2}\Big[W-\frac{\partial^{\mu}Z_{\mu}}{m}\Big]^{2}+\frac{1}{2}B^{[\mu\alpha][\nu\beta]}R^{(1)}_{\mu\alpha\nu\beta}(W-me)\Bigg\}\;,\label{sibr}\eea

\no where $R^{(1)}_{\mu\alpha\nu\beta}$ is the linearized version of the Riemann curvature tensor under a weak field expansion $g_{\mu\nu}=\eta_{\mu\nu}+e_{(\mu\nu)}$, namely,

\bea R^{(1)}_{\mu\alpha\nu\beta}(e)=\frac{1}{2}(\partial_{\mu}\partial_{\beta}e_{(\alpha\nu)}-\partial_{\mu}\partial_{\nu}e_{(\alpha\beta)}-\partial_{\alpha}\partial_{\beta}e_{(\mu\nu)}+\partial_{\alpha}\partial_{\nu}e_{(\mu\beta)}) \;. \eea

Similar to the passage from the intermediate action (\ref{BF-1o}) to the Cremmer-Scherk model (\ref{BF}), the integration over $W_{\mu\nu}$ in (\ref{sibr}) results in the fourth-order BR model \cite{ks}

\bea S^{(4)}_{BR}&=&\int{d^{D}x}\Bigg\{-\frac{1}{2}\partial_{\mu}e_{(\alpha\beta)}\partial^{\mu}e^{(\alpha\beta)}+\frac{1}{2}\partial_{\mu}e\partial^{\mu}e-\partial_{\mu}e\partial_{\nu}e^{\mu\nu}+\partial_{\mu}e^{(\mu\nu)}\partial^{\alpha}e_{(\alpha\nu)}\nn\\
&&\qquad-\frac{m}{2}B^{[\mu\alpha][\nu\beta]}R^{(1)}_{\mu\alpha\nu\beta}(e)+\frac{1}{2}(\partial_{\alpha}\partial_{\beta}B^{[\alpha\mu][\beta\nu]})^{2}-\frac{1}{2}\frac{(\partial_{\alpha}\partial_{\beta}B^{\alpha\beta})^{2}}{(D-1)}\Bigg\} \;.\label{BR}\eea

\no The model (\ref{BR}) is invariant under transformations $\delta_{\Phi}B_{[\mu\alpha][\nu\beta]}$ which leave $\p^{\mu}\p^{\nu}B_{[\mu\alpha][\nu\beta]}$ invariant and under linearized reparameterizations $ \delta_{\xi}{e}_{(\mu\nu)}=\partial_{\mu}\xi_{\nu}+\partial_{\nu}\xi_{\mu}$.

The equations of motions of the action (\ref{BR}) are

\bea G^{(1)}_{\mu\nu}(e)=\frac{1}{2}\partial^{\alpha}\partial^{\beta}B_{[\mu\alpha][\nu\beta]} \;,\label{eq-1-BR}\eea

\bea R^{(1)}_{\mu\alpha\nu\beta}(e-\mathbb{B})=0 \;,\label{eq-2-BR}\eea

\no where $G^{(1)}_{\mu\nu}$ is the linearized Einstein tensor, $G^{(1)}_{\mu\nu}(e)=R^{(1)}_{\mu\nu}(e)-(1/2)\eta_{\mu\nu}R^{(1)}(e)$,
and we have defined the symmetric tensor

\bea \mathbb{B}_{\mu\nu}=-\,\frac{1}{m^{2}}\Big[\partial^{\alpha}\partial^{\beta}B_{[\mu\alpha][\nu\beta]}-\frac{1}{(D-1)}\eta_{\mu\nu}\partial^{\alpha}\partial^{\beta}B_{\alpha\beta}\Big] \;.\label{eq-3-BR}\eea

\no Using those results in (\ref{eq-1-BR}) we have

\bea G^{(1)}_{\mu\nu}(\mathbb{B})=-\,\frac{m^{2}}{2}(\mathbb{B}_{\mu\nu}-\eta_{\mu\nu}\mathbb{B}) \;.\eea

\no Those are the Fierz-Pauli equation of motion for the field $\mathbb{B}_{\mu\nu}$, from which  we can deduce the Klein-Gordon equation and the Fierz-Pauli constraints

\bea (\square-m^{2})\mathbb{B}_{\mu\nu}=0\quad,\quad \partial^{\mu}\mathbb{B}_{\mu\nu}=0\quad,\quad \mathbb{B}=\eta^{\mu\nu}\mathbb{B}_{\mu\nu}=0\;.\eea

\no The dual mapping between the Fierz-Pauli and the BR models is

\bea e_{(\mu\nu)}\;\leftrightarrow\;\mathbb{B}_{\mu\nu} \;.\eea

\no In the special case of $D=2+1$ dimensions we can rewrite the tensor field $B^{[\mu\alpha][\nu\beta]}$ in terms of a symmetric rank-2 tensor $\ell_{\lambda\sigma}$:

\bea B^{[\mu\alpha][\nu\beta]}\equiv\epsilon^{\mu\alpha\lambda}\epsilon^{\nu\beta\sigma}\ell_{\lambda\sigma} \;.\eea

\no Such that the BR model (\ref{BR}) can be written as\footnote{In $D=2+1$ the Weyl tensor $C_{\mu\alpha\nu\beta}$ is zero identically, so that the Riemann curvature tensor can be written in terms of the Einstein tensor, $R_{\mu\alpha\nu\beta}=\epsilon_{\mu\alpha\lambda}\epsilon_{\nu\beta\sigma}G^{\lambda\sigma}$.}

\bea S^{(D=3)}_{BR}&=&2\int{d^{3}x}\Big\{\mathcal{L}_{EHL}(h)-\ell^{\mu\nu}G^{(1)}_{\mu\nu}(h)+\frac{1}{m^{2}}\mathcal{L}_{K}({\ell})\Big\} \;,\label{BG}\eea

\no where  $h_{\mu\nu}\equiv e_{(\mu\nu)}$ and

\bea \mathcal{L}_{LEH}(h)&=&-\frac{1}{4}\partial_{\mu}h_{\alpha\beta}\partial^{\mu}h^{\alpha\beta}+\frac{1}{4}\partial_{\mu}h\partial^{\mu}h-\frac{1}{2}\partial_{\mu}h\partial_{\nu}h^{\mu\nu}+\frac{1}{2}\partial_{\mu}h^{\mu\nu}\partial^{\alpha}h_{\alpha\nu} \;,\\
\mathcal{L}_{K}(\ell)&=&\frac{1}{4}(\square\ell_{\mu\nu})^{2}+\frac{1}{2}\square(\partial_{\mu}\ell^{\mu\nu})^{2}+\frac{1}{8}(\partial_{\mu}\partial_{\nu}\ell^{\mu\nu})^{2}+\frac{1}{4}\square\ell_{\mu\nu}\partial^{\mu}\partial^{\nu}\ell-\frac{1}{8}(\square\ell)^{2} \;.\eea

The Lagrangian $\mathcal{L}_{LEH}(h)$ is the linearized version of the Einstein-Hilbert Lagrangian $\sqrt{-g}R(g)$ in the weak field expansion $g_{\mu\nu}=\eta_{\mu\nu}+h_{\mu\nu}$, while $\mathcal{L}_{K}(\ell)$ is the linearized version of the K-term of the New Massive Gravity \cite{bht}

\bea K(f)=R^{2}_{\mu\nu}(f)-\frac{3}{8}R^{2}(f) \;,\eea

\no in the weak field expansion $f_{\mu\nu}=\eta_{\mu\nu}+\ell_{\mu\nu}$.

So the BR model in $D=2+1$ can be interpreted as a linearized version of a bimetric model,

\bea S_{bim}[g,f]=2\int{d^{3}x}\sqrt{-g}\Big[\,\frac{7}{2}R(g)+f^{\mu\nu}G_{\mu\nu}(g)+\frac{K(f)}{m^{2}}\Big] \;,\label{bim}\eea

\no where $f^{\mu\nu}f_{\nu\beta} = \delta^{\mu}_{\beta} $. In the weak field expansion of the metrics $g_{\mu\nu}$ and $f_{\mu\nu}$ we have until the quadratic order
\bea \sqrt{-g}R(g)\;&\rightarrow&\;\mathcal{L}_{LEH}(h)\;,\label{leh}\\
\sqrt{-g}f^{\mu\nu}G_{\mu\nu}(g)\;&\rightarrow&\;-\ell^{\mu\nu}G^{(1)}_{\mu\nu}(h)-\frac{5}{2}\mathcal{L}_{LEH}(h) \;,\label{fg}\\
\sqrt{-g}K(f)\;&\rightarrow&\;\mathcal{L}_{K}(\ell) \;.\label{K}\eea

\no Where on the right hand side of (\ref{leh})-(\ref{K}) the indices are raised and lowered with the flat metric $\eta_{\mu\nu}$.
The unusual coefficient $7/2$ in the Einstein-Hilbert term can be explained by the extra term $-5/2$ in (\ref{fg}). The bimetric model (\ref{bim}) belongs to a larger class of bimetric models which use different types of derivative couplings between the metrics and describe massive spin-2 particles in $D=2+1$. We are currently investigating such new class of models.

\section{Conclusion}

The action $S_{P-dual}^{(1)}$, see (\ref{Tens-s1-1a}), is a dual version of the usual first order formulation of the Maxwell theory. Its equations of motion (in vacuum ) become explicitly eletric-magnetic duality invariant, see (\ref{dualF}). We believe that $S_{P-dual}^{(1)}$ can be generalized for arbitrary integer spin as a dual version of the Vasilev's  parent action \cite{Vasiliev}. Starting with its spin-2 version $S_{P-dual}^{(2)}$, which has appeared before in \cite{bch},  we have performed a Kaluza-Klein dimensional reduction from $D+1$ to $D$ dimensions and obtained  a parent action for massive spin-2 particles in $D$-dimensions, see (\ref{Tens-s2-1a}), which provides a unified framework for obtaining rather different massive spin-2 models.

We have derived five dual models, besides the Fierz-Pauli \cite{fp}, Curtright-Freund \cite{Curt} and
Cassini-Montemayor-Urrutia \cite{cmu} models, we have also obtained, to the best we know, a new model, see (\ref{BT}), and the more recent ``BR'' model of \cite{ks} which is a kind of spin-2 generalization of the spin-1  topologically (in $D=4$) massive BF model or Cremmer-Scherk model \cite{Cremmer}. In the BR model the linearized Riemann tensor $R^{(1)}_{\mu\nu\alpha\beta}(h)$ replaces the spin-1 curvature $F_{\mu\nu}(A)$. The model describes massive spin-2 particles  without breaking the linearized diffeomorphism invariance and without explicit Stueckelberg fields either. Differently from \cite{ks}, where the Riemann tensor is an input, here it  is an output that emerges from a solution of a functional constraint. This fact  may be relevant in the investigation of higher spin ($s\ge 3$)  geometries. The BR model has inspired us to introduce a new bimetric model describing massive gravitons in $D=2+1$. It is one specific example of a multiparametric family of new bimetric models in $D=2+1$ that we are currently investigating beyond the linearized approximation  \cite{dbs}.

 The new model (\ref{BT})  is also invariant under linearized diffeomorphisms even without explicit Stueckelberg fields akin to the BR model but the fields have a different tensor structure. It contains a mixed symmetry tensor, whose kinetic term is the same one of the Curtright-Freund model, and the rank-2 tensor must be  non symmetric. The two models are   similar to the spin-1 Cremmer-Scherk model but the BR model,  after the elimination of Stueckelberg fields, has a smoother massless limit since the kinetic term for the field
 $B_{[\mu\nu][\alpha\beta]}$ describes massless spin-1 particles, see \cite{Siegel}, while the kinetic term
for the field $B_{\mu[\alpha\beta]}$ in (\ref{BT}) has no particle content. However, if we keep the Stueckelberg fields the new model (\ref{BT}) has a smooth massless limit preserving the same number of degrees of freedom of the massive case.  Both models (\ref{BT}) and (\ref{BR}) may be used as an alternative starting point for building up diffeomorphism invariant massive gravity models.

It is remarkable that the antisymmetric components $e_{[\mu\nu]}$ play a crucial role in the derivation of the BR model and of the new model (\ref{BT}) while it plays no role on the Fierz-Pauli side. The key role of $e_{[\mu\nu]}$ in proving duality between the Curtright-Freund and the Fierz-Pauli models has been pointed out in \cite{kmu,gkmu}.

We are currently investigating also higher spin and curved background generalizations of
the dual action (\ref{Tens-s2-1a}).  The curved space generalization may be relevant for recent discussions \cite{hinter1,hinter2} on electric/magnetic duality \cite{deser-em} which appears in  partially-massless theories  in $D=3+1$. In particular, it may be useful for the search of an electric/magnetic
 duality invariant action still
covariant with respect to the curved background reparametrizations.

\section{Acknowledgements}

The work of D.D. is partially supported by CNPq  (grant 306380/2017-0) while A.L.R.dos S. has been supported by CNPq (grant 160784/2019-0). A.L.R.dos S. thanks INCT-FNA Programm.

\end{document}